\documentclass[doublecol]{epl2} 

\usepackage{graphicx}% Include figure files
\usepackage{dcolumn}% Align table columns on the decimal point
\usepackage{bm}% bold math
\usepackage{hyperref}
\usepackage{xcolor}
\usepackage{amssymb}
\usepackage{amsmath}
\usepackage{float}
\usepackage{caption}
\usepackage{multicol}

\title{Effective single particle theory for active particles using local density fluctuations}
\shorttitle{Effective single particle theory for active particles} %Insert here a short version of the title if it exceeds 70 characters

\author{Jayam Joshi\inst{1} \and Pawan Kumar Mishra\inst{1} \and Shradha Mishra\inst{1}}
\shortauthor{Jayam \etal}

\institute{                    
  \inst{1} Department of Physics, Indian Institute of Technology (BHU) Varanasi\\
}

\abstract{
We characterize the dynamic non-equilibrium steady state behavior of active particles using density fluctuations in the system. We analyze the effective local density around a particle in the steady state and numerically calculate its mean, variance and auto-correlation. Thus, using local density and its statistical properties as a temporally correlated stochastic variable, we develop an effective single-particle theoretical model and analytically derive an  expression for the particle's diffusivity as a function of the global packing density in the system. Our theory accurately predicts the transport properties of an active particle, validated against numerical simulations. Unlike mean-field theory, which fails at high packing densities due to significant density fluctuations from dynamic cluster formation, our model remains effective across all densities. It also captures the well-known phase transition beyond a critical packing density. The key novelty of our model lies in the introduction of a stochastic local density field, which encapsulates the effect of steric interactions on an active particle and helps predict single-particle behavior in a collection—a feature often absent in standard active matter models. This approach could be useful in experimental setups where fluctuations in local density around a tagged particle are measurable.}

\begin{document}

\maketitle

%%%%%%%%%%%%%%%%Introduction%%%%%%%%%%%%%%%%%%%%%%%%%%%%%%%%%%%%%%%%%%%%
\textit{Introduction-} Active matter constitutes a distinct class of non-equilibrium systems that are composed of particles that consume energy from their surroundings and convert it into motion or exert mechanical forces\cite{Toner1995,tonertupre1998,rmp2013Marchetti,rmp2016Bechinger,pr2012Vicsek,TopicalReview2020}. It encompasses a wide range of examples, from the biological realm, including microswimmers such as E.Coli bacteria \cite{bacteria}, collective behaviors observed in groups of animals such as fish schools \cite{fish} and bird flocks \cite{birds}, as well as artificial microswimmers, colloidal suspensions, and robots. On an individual particle level, because of the constant flux of energy, an active particle is in a perpetual far-from-equilibrium state. In such a non-equilibrium state, the particle produces net entropy while breaking the usual equilibrium conditions of detailed balance and time-reversal symmetry. Thus, activity of a single particle adds complexity in its motion, and hence its motion is studied through frameworks of non-equilibrium statistical physics. \\
Another layer of complexity in behavior of active matter systems is added when active particles are in a collection, and the individually out-of-equilibrium particles interact to display emergent steady-state phenomena like novel phase transitions, long-range ordering, giant number fluctuations, collective motion and self-organization, which are not observed in equilibrium systems \cite{active_1, mips_3, mips_4, mips_5,  mips_7, mips_8, mips_9}. \\ 
There is a gap in understanding the dynamic behavior of active matter systems in non-equilibrium steady states specific to its effect on a single particle. In this study, we take a top-down approach, first characterizing the steady-state behavior of the system using a stochastic local density variable in numerical simulations, and then using it to theoretically calculate the effective dynamics of a single particle. We use active Brownian particles \cite{abp_1,abp_2,abp_3,abp_4,ABP,dolai2018phase,semwal2024macro}, which are self-propelled disk-shaped particles that move at constant speed, maintaining their direction for some time before reorientation due to thermal fluctuations. This system exhibits motility-induced phase separation (MIPS) above a threshold packing density, characterized by the dynamic formation and dissolution of macroscopic clusters.
We theoretically model dynamic cluster formation and dissolution by introducing a local density variable. Starting with an analysis of local density fluctuations around a tagged particle in simulations, we proceed to calculate effective local density fluctuations by averaging over all particles and ensembles. In the steady state, we numerically compute key statistical properties, such as mean, variance, and temporal auto-correlation of local density, which remain constant, characterizing it as a stationary stochastic variable in time. In this state, the local density fluctuates with a constant variance around a stable mean value, while its temporal auto-correlation decays exponentially. This behavior reveals a finite correlation time for the persistence of local density around a particle, akin to an Ornstein-Uhlenbeck process in the stationary regime \cite{OU}. Both the mean local density and correlation time increase with global packing density along with a crossover in behavior near a critical density, consistent with motility-induced phase separation (MIPS)\cite{wang2014bulk, schmittmann1998driven, morse2021direct, kramar2022intermittency}. \\
We then formulate the effective Langevin equations for a single particle, incorporating the effects of interactions through the local density. Solving these equations analytically, we derive an exact expression for particle diffusivity, which closely matches the numerical results for all global packing densities considered. Remarkably, our effective single-particle theory improves on mean-field theory by accounting for density fluctuations in the phase-separated state, yielding accurate diffusivity estimates in both dilute systems and dense systems.
\begin{figure}[H]
\begin{minipage}{0.49\textwidth}
    \includegraphics[width=1.0\linewidth, height=10.0cm, keepaspectratio]{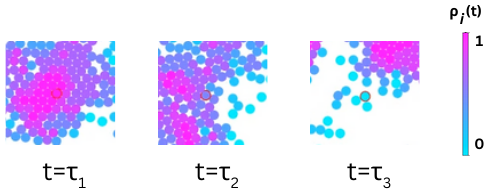}
    \caption{Zoomed-in snapshot of a part of the system at three random time instants, depicting the local density around each particle using the attached color bar. By tracking the local density $\rho_i$ around a fixed particle (tagged by red boundary) at all times, we obtain the local density stochastic time series $\rho_i (t) $.}
    \label{fig:rho_i}
\end{minipage}
\end{figure} 
\textit{Model-} We aim to derive effective Langevin equations for a single particle, incorporating effects of interactions in the non-equilibrium steady state. Guided by simulations of active Brownian particles and prior studies \cite{mips_5}, we assume that steric hindrance reduces the effective particle speed. Thus, we make an ad-hoc hypothesis of a local density-dependent speed and formulate the corresponding Langevin equations:
\begin{equation}
	\partial_t{\bf{r}}=v_{\rho} {\bf \hat{{n}}}
 \label{eqn_pos}
\end{equation}
\begin{equation}
	 \partial_t\theta= \sqrt{2D_{R}}  \eta (t)
   \label{eqn_theta}
\end{equation}
here, the self propelling speed $v_{\rho}$ of the particle is dependent on the time-dependent local density $\rho (t)$. ${\bf \hat{n}}(t) = (cos\theta(t), sin\theta(t))$ is the orientation vector of the particle, pointing in the direction of its velocity which makes angle $\theta(t)$ with the x-axis. $D_R$ is the rotational diffusion constant, and $\eta(t)$ is white Gaussian noise which has zero mean and is delta correlated $ \langle \eta(t_1)\eta(t_2) \rangle = \delta(t_1 - t_2)$. We comment on the exact form of dependence of speed of particle on local density later in the article, but for now we analyze the introduced local density variable using extensive simulations of a system of $N$ sterically interacting active Brownian particles. In the numerical simulations, we quantify the local density near a particle as the number of particles it is contact with normalized by six, assuming the formation of a hexagonal close-packed structure (HCP) for strong clustering.\\ 
\begin{figure}[H]
\begin{minipage}{0.49\textwidth}
    \centering
    \includegraphics[width=\linewidth, height=12.5cm, keepaspectratio]{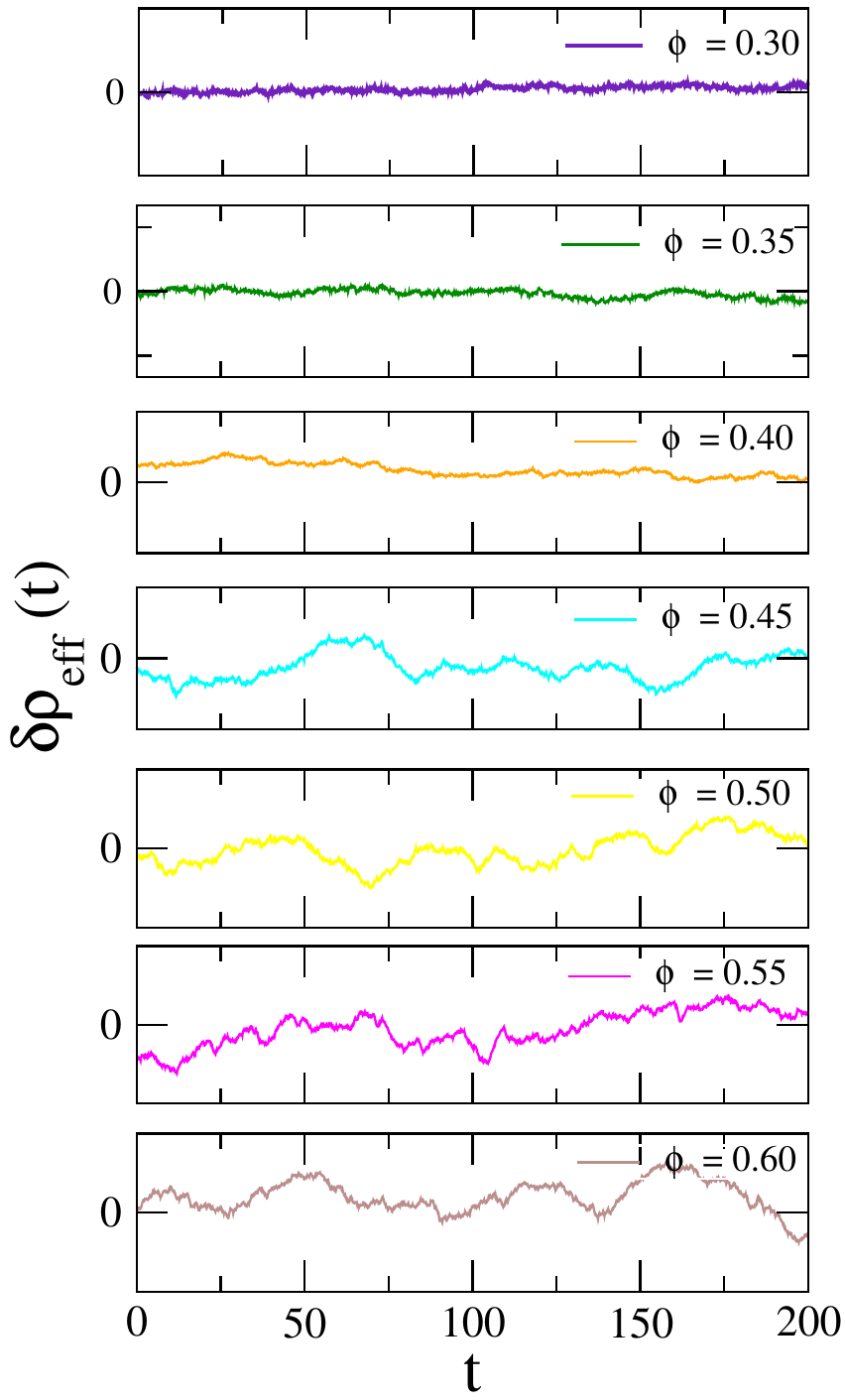}
    \caption{ Effective local density fluctuations (\textit{Y-axis limits} $\epsilon[ -0.15,0.15]$) as a function of time in the steady state of the system, averaged over all particles and ensembles and calculated for a range of system packing densities. The activity of particles and system size is fixed at $v_0 = 1.0$ and $N=2000$.}
    \label{fig:rho_eff_t}
\end{minipage}
\end{figure} 
{\em Numerical simulation:}- We perform numerical simulations of a system consisting of $N$ disc-shaped active Brownian particles (ABPs) of radius $a$ in a box of dimension $L$ with periodic boundary conditions in both directions. We simulated the system by numerically integrating the overdamped Langevin equations of motion for the position $\bm{r}_i$ and orientation $\theta_i$ of each ABP given by: 
\begin{equation}
	\partial_t{{\bf{r}}_i}=v_0 {\bf \hat{{n}}}_i + \mu \sum_j^{} \bm{F}_{ij}
 \label{num_sim_pos}
\end{equation}
\begin{equation}
	 \partial_t\theta_i = \sqrt{2D_{R}}  \eta_i (t)
   \label{num_sim_theta}
\end{equation}
$v_0$ is the self-propulsion of each particle. The ABPs interact according to a short-ranged and repulsive harmonic potential such that the force $\bm{F}_{ij}$ between particles $i$ and $j$ is given by: $\bm{F}_{ij} = -k(2a - {r}_{ij})\bm{\hat{r}}_{ij}$ if $r_{ij} \leq 2a $ and $\bm{F}_{ij} = 0$ if $r_{ij} > 2a $. Here, $\bm{r}_{ij}$ is the vector from the $j^{th}$ to the $i^{th}$ particle. The steric force constant $k$ and mobility $\mu$ set the elastic time scale $(\mu k)^{-1}$ in the system. The velocity orientation angle $\theta_i$ of the $i^{th}$ particle undergoes a Brownian motion governed by white Gaussian noise $\eta_i (t)$ which has zero mean and is delta correlated such that $\langle \eta_i(t_1)\eta_j(t_2) \rangle = \delta_{ij}\delta(t_1 - t_2)$.  The rotational time scale ${D_R}^{-1} = 10^{4} \Delta t$. The small integration time step $\Delta t = 0.001$.
We start with random initial  positions and orientations of ABPs and simulate for total simulation steps $T_{steps}=5 \times 10^6$. The actual time $t = \Delta t \times T_{steps}$.  Averaging is performed over $N_{ensembles}=20$ independent realizations for improved statistics. In numerical analysis, we have used a set of different values of global packing densities $\phi = \pi a^{2} N/L^2$ (ranging from $0.1$ to $0.65$) which is varied by changing the box dimension $L$. The activity/Peclet number of the system is varied by self-propulsion speeds $v_0 = (0.75, 1.0, 1.25, 2.0$ and $3.0$). To perform  finite system size analysis, we have also performed simulations for different system sizes by varying the number of particles $N = (300,500,600,800,1000,2000)$. \\
Using a density-tracking algorithm in the microscopic simulations mentioned above, the local density near a particle defined by number of contacting particles normalized by six, is tracked for each particle and for a fixed period of time in the steady state. In Fig. \ref{fig:rho_i}, local density $\rho (t)$ of each particle is shown by a color bar, using a zoomed in snapshot of the simulation at three random time instants. $\rho (t)$ is a stochastic variable in time and its value fluctuates randomly in a set of values between 0 to 1. To better understand single-particle behavior using the stochastic time-dependent local density around it, we calculate the effective local density fluctuations $\rho (t)$ for a time   $t=200 sec$ in the steady state by averaging it over all particles in the system and simulation ensembles to obtain $\rho_{eff}(t)$. Then, we analyze its fluctuation $\delta \rho_{eff} (t)$ from the mean local density in the system during the steady state $\rho_0 = \langle\rho_{eff} (t) \rangle_t$. By doing this, we aim to investigate and visualize the average density fluctuations around a single active particle in the presence of interactions with other active particles in the collection. We repeat this analysis for the entire range of packing densities, activity and system sizes considered. From here onward we simply omit writing $\rho_{eff}$ and use local density $\rho$ in place of effective local density everywhere for convenience. Fig. \ref{fig:rho_eff_t} shows the effective local density fluctuations in a system with activity $v_0 = 1.0$ and system size $N = 2000$. The local density fluctuates around a constant value in the steady state for all global packing densities, but the fluctuations are very small for low global packing densities. Hence, in the dilute or low packing density regime, we can assume that the active particle is effectively navigating in an environment of homogeneous density $\rho_0$. Further, since density fluctuations are negligible in the dilute limit, we can approximate the local density around it as a mean field and its effective self propulsion speed varying as $v_{\rho} = v_0 (1 - \lambda \rho_0)$, where $\lambda$ is an appropriate fitting parameter calculated from numerical results \cite{active_1} and $\rho_0 = \langle \rho_{eff}(t)\rangle$. The mean-field theory accurately estimates particle diffusivity in the dilute limit. However, beyond a critical packing density, the local density shows significant fluctuations with a non-zero variance and thus the mean field approximation fails to capture the dynamics of the particle. We aim to capture the effect of density fluctuations and theoretically calculate the effective diffusivity of a particle in the entire range of packing densities from the dilute to dense.
\begin{figure}[H]
\begin{minipage}{0.49\textwidth}
\centering
    \includegraphics[width=0.9\linewidth, height=8.0cm, keepaspectratio]{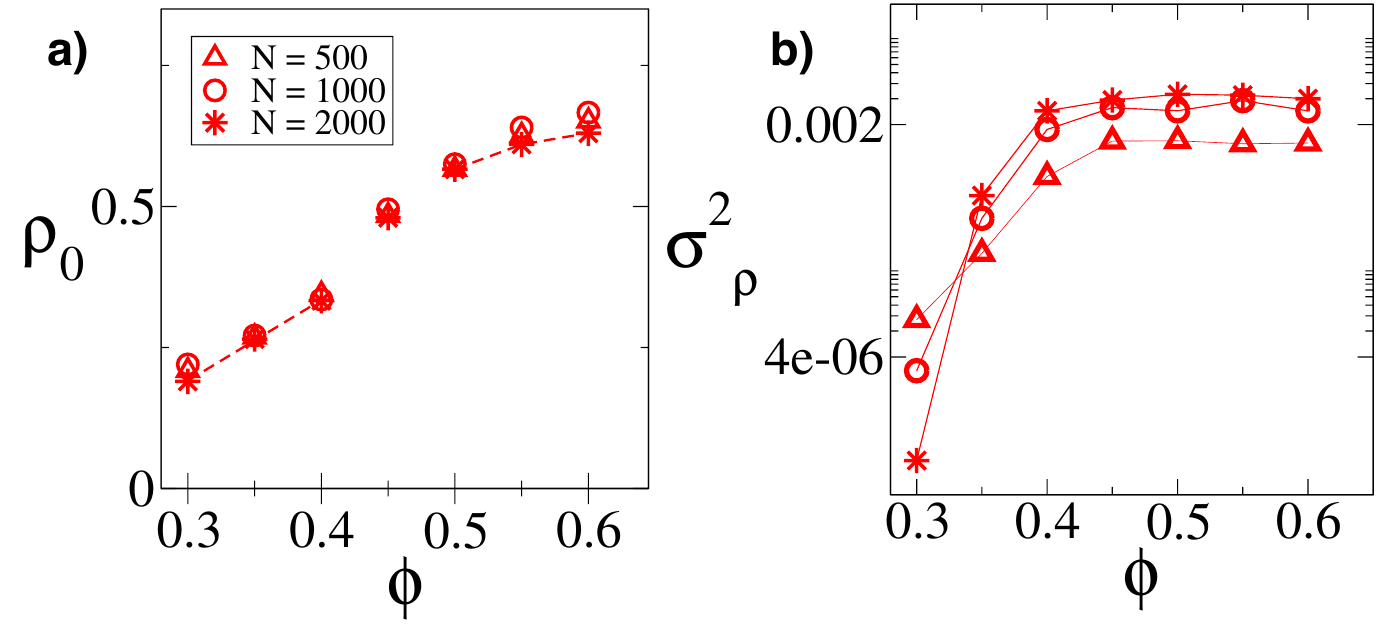}
    \caption{a) Mean value of effective local density around a particle in the steady state, as a function of the system packing density $\phi$ and varying system sizes $N$. b) Variance of local density fluctuations as a function of system packing density. The activity of particles is fixed to $v_0 = 1.0$. The lines are to simply join the data points.}
    \label{rho_mean_var}
\end{minipage}
\end{figure} 
For further insight into the nature of the local density variable, we calculate the mean $\rho_0$ and variance ${\sigma_{\rho}}^2$ of the local density variable for fixed activity of particles $v_0 = 1.0$ and varying system packing densities $\phi$ and finite system sizes $N$. We find that the mean and variance of fluctuations are time independent in the steady state. In Fig. \ref{rho_mean_var} a), we plot the mean local density around a particle in the system, by time-averaging $\rho(t) $ in the steady state, $\rho_0 = \langle \rho(t) \rangle_t$. This is the mean value around which the local density fluctuates. $\rho_0$ increases monotonically and approximately linearly with global packing density. Finite-system size analysis (with N = 500, 1000, and 2000) shows the mean local density around a particle remains roughly constant with increasing system size. It saturates at large packing densities due to geometric effects. The linear variation shows a crossover in behavior for intermediate $\phi$, supporting the existence of motility-induced phase separation into a dense dynamic cluster state and a surrounding vapor.

\begin{figure}[H]
\begin{minipage}{0.49\textwidth}
\centering
    \includegraphics[width=\linewidth, height=9.75cm, keepaspectratio]{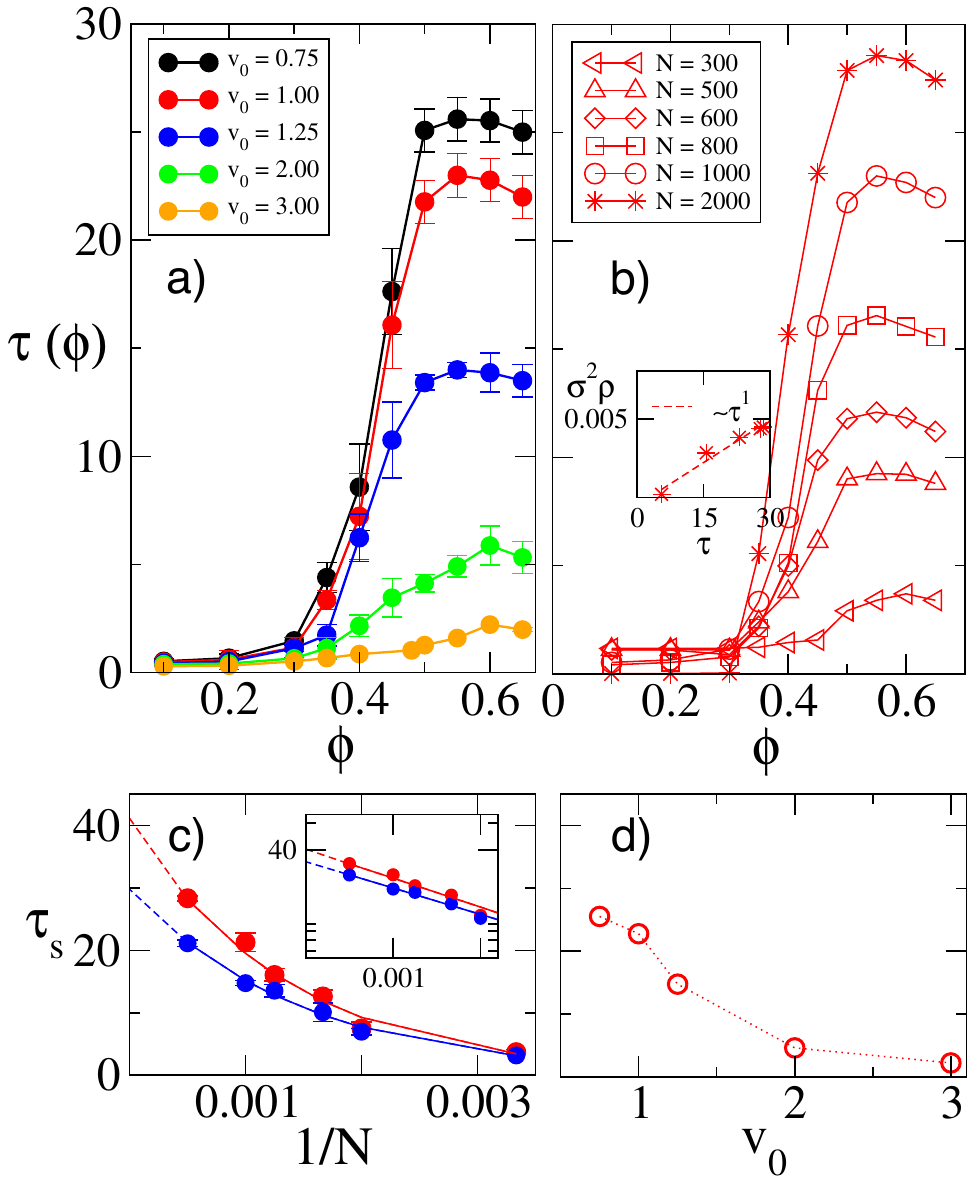}
    \caption{(a) Shows the plot of correlation time $\tau$ vs.  system packing density $\phi$ of system for five different values of self-propulsion speed $v_0=0.75,1.00,1.25,2.00,3.00$. (b) Variation of $\tau(\phi)$ for different system sizes N for fixed activity $v_0 = 1.0$. Inset: linear relation between variance and correlation time of density fluctuations for large system$(N=2000)$(c) Shows the variation of saturated correlation time $(\tau_s)$ at packing density $\phi_o = 0.6$ vs. system size $(1/N)$ for two speeds $v_0=1.0$ and $v_0=1.25$, exponential fits to the data and extrapolation of the fits to obtain finite $\tau_{\infty}$ for bulk behavior. Inset: same data and curves on semi-log-y scale. (d) The variation of  $\tau_s$ vs. $v_0$. The line is simply to join the points.}
    \label{corr_time}
\end{minipage}
\end{figure}

In Fig. \ref{rho_mean_var}b, we compute the variance $\sigma^2_{\rho}$ to quantify fluctuations in local density around the mean. As discussed earlier, the variance is negligible at low global packing densities, where the mean-field approximation holds. It increases monotonically to a nonzero saturated value at high global packing densities. Finite-size analysis of the system shows that for $\phi>\phi_c$, $\sigma^2_{\rho}$ increases with system size $N$, while for $\phi <\phi_c$, it decreases to $0$. Above the critical density $\phi_c$, the system undergoes motility-induced phase separation, leading to the coexistence of dense clusters and dilute vapor regions; larger systems permit the development of more pronounced density fluctuations between these phases. In contrast, below $\phi_c$, the system remains homogeneous, and increasing the system size leads to enhanced averaging over local fluctuations, thus reducing the variance. We aim to account for these fluctuations to calculate the diffusivity of an active particle.
\begin{figure}[H]
\begin{minipage}{0.49\textwidth}
\centering
    \includegraphics[width=\linewidth, height=15.0cm, keepaspectratio]{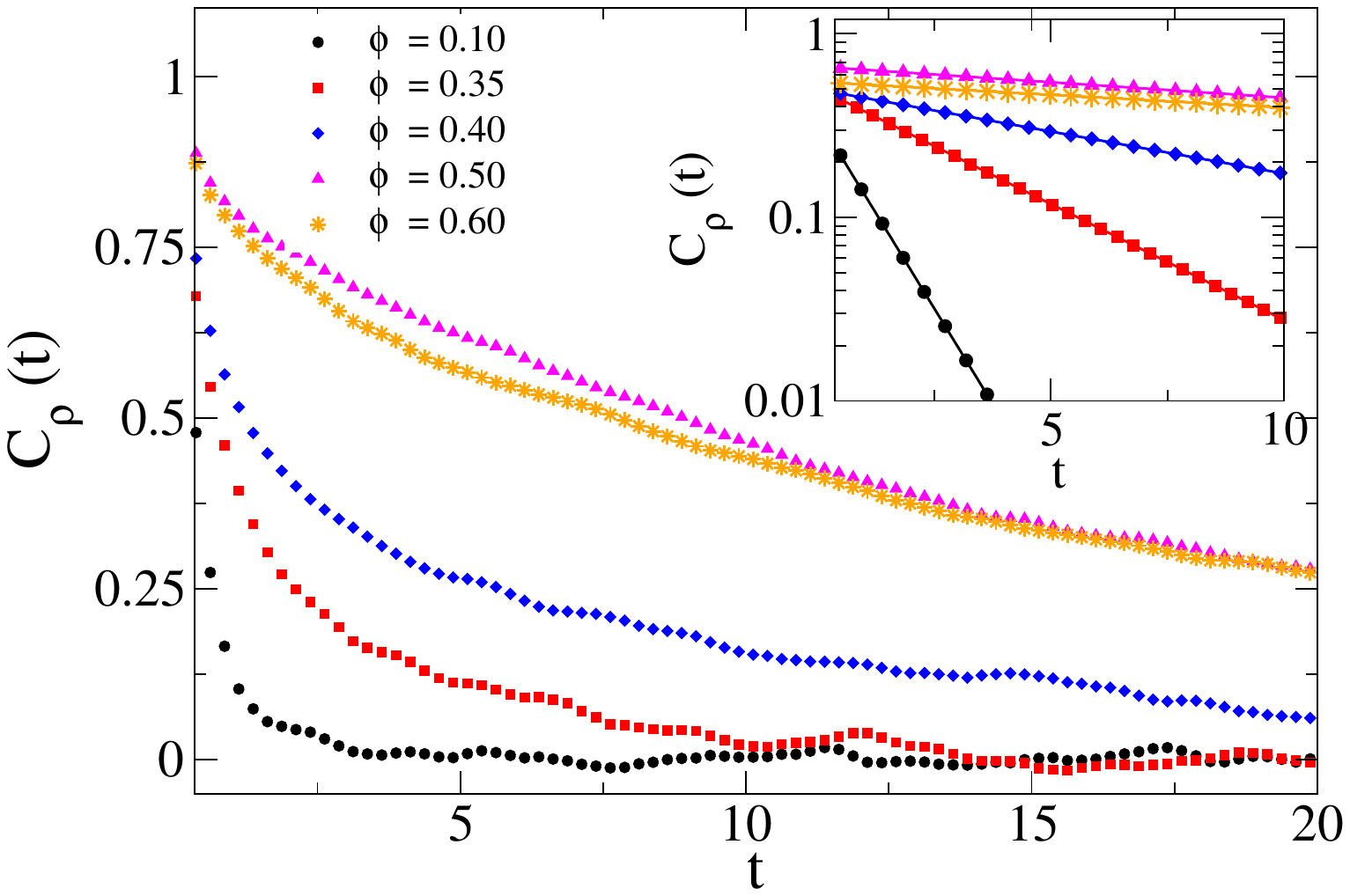}
    \caption{Local density auto-correlation $\mathcal{C}_{\rho}(t)$ vs. time for different system packing density. $\phi$ in a system of ABP with activity $v_0 = 1.0$ and system size $N = 2000$. Inset: $\mathcal{C}_{\rho}(t)$ zoomed for short times and plotted on semi-$\log$-y scale. The straight lines are exponential fits to the data from which correlation time is calculated.}
    \label{density}
\end{minipage}
\end{figure} 
We further analyze another property of local density fluctuations by calculating its temporal auto-correlation $ \mathcal{C}_{\rho}(t) = \frac{\langle\delta \rho_i(t_0) \delta \rho_i(t_0+t)\rangle}{\sigma^2_\rho}$, where $\langle..\rangle$ means, averaging over all the particles and reference times from $t=t_0$ to $t=t_0 + 100$, once the system is in the steady state at $t=t_0$. $\delta \rho_i(t)$ is the fluctuation in local density of $i^{th}$ particle from the mean density $\rho_0$ of the system. In Fig.\ref{density} we show the plot of normalized $\mathcal{C}_{\rho}(t)$ vs. $t$ for different packing densities $\phi$, and it clearly suggests an exponential decay with time.  This physically signifies that the local density around a particle tends to persist for a finite correlation time after which it decays, due to the dynamic formation and dissolution of clusters in the high packing density regime of the system. The density correlation time $\tau$ is obtained from exponential fits of the form $\mathcal{C}_\rho (t) = exp(-t/\tau)$ as shown in the inset of Fig. \ref{density}. \\
In Fig. \ref{corr_time} (a) $\tau$ is negligible for $\phi < 0.3$, as in this dilute regime, only a homogeneous micro-clustered phase exists, with clusters rapidly forming and breaking. As packing density increases, $\tau$ grows monotonically, shows a discontinuous jump near intermediate $\phi_c$, and saturates to a non-zero value at high densities. This saturation indicates the formation of macroscopic clusters at the onset of MIPS.
The finite-system size analysis in Fig. \ref{corr_time}(b) shows that for $\phi>\phi_c$, $\tau$ increases with the system size $N$, since larger systems stabilize dense macroscopic clusters in the MIPS state. In contrast for $\phi < \phi_c$, $\tau$ decreases with $N$, reflecting enhanced averaging over the rapidly fluctuating micro-clusters in the homogeneous phase. The sharpening discontinuous jump with increasing $N$ further highlights the first-order-like characteristics of the phase transition.
We also performed finite size analysis of the saturated correlation time $\tau_s$ in dense system with $\phi = 0.6$ for two different activity of particles $v_0 = 1.0$ and $1.25$. Fig.  \ref{corr_time}(c) shows  $\tau_s$ {\em vs.} $1/N$, data points are obtained from numerical simulation and the curves are non-linear fits to the data. Extrapolation of the fits meet the y-axis at a finite value and suggest that a finite value of ${\tau_s}_{\infty}$ exists for bulk behavior corresponding to $N  \rightarrow \infty $ (shown by the dashed line). Additionally, the non-linear fit is well approximated by an exponential function $ \tau_s = {\tau_s}_{\infty} exp( - \frac{\alpha}{N})$, and plotting $\log (\tau_s)$ vs $1/N$ on the semi-log-y scale (inset of Fig. \ref{corr_time} (c) confirms this, as the straight-line behavior intersects the y-axis at a finite value, reinforcing that $\tau_s$ saturates for very large system sizes. We also calculated the auto-correlation time $\tau$ for other self-propulsion speeds $v_0 = 0.75$, $1.25$, $2.0$ and $3.0$. For all the cases, the behavior of $\tau$ remains the same. Higher activity makes the motion of the particles more persistent which increases the dynamic nature of the clusters; hence, $\tau$ decreases.  Fig. \ref{corr_time}(d) shows the dependence  $\tau_s$ on the activity $v_0$ for $\phi=0.6$. $\tau_s$ decays monotonically with increasing activity. Hence  the enhanced motility makes the cluster unstable. Additionally, the inset of Fig. \ref{corr_time} (b) shows that for bulk behavior, $\sigma^2_{\rho}$ is directly related to and varies linearly with $\tau$. So the effective local density fluctuations or $\sigma^2_{\rho}$ also assume a finite value for very large system size $N \rightarrow \infty$. This suggests that the effective local density around a particle in a collection of active particles can be modeled as a temporally correlated stationary stochastic variable  similar to colored noise.

Consequently, we can state the properties of local density derived from numerical simulations:
\begin{equation}
    \langle \rho(t) \rangle  = \rho_0 
    \label{rho_mean}
\end{equation}
\begin{equation}
    \mathcal{C}_{\rho}(t) = \langle \delta \rho(0)\delta \rho(t) \rangle  = \sigma^2_\rho \exp(-t/\tau)
    \label{rho_corr}
 \end{equation}
Using local density fluctuations and its properties as the main ingredient, we proceed to calculate the effective dynamics of a single particle.\\
%%%%%%%%%%%%%%%%%%%%%%%%%%%%%%%%%%%%%%%%%%%%%%%%%%%%%%%%%%%%%%%%%%%%%%%%%%%%%%%%
%%%%%%%%%%%%%%%%%%%%%%%Analytical Theory%%%%%%%%%%%%%%%%%%%%%%%%%%%%%%%%%%%%%%%%%
 \textit{Analytical Theory-} Now, we recall the effective Langevin equations of a single particle formulated earlier in \ref{eqn_pos} and \ref{eqn_theta}, and solve them by assuming a specific functional dependence of particle speed on local density around it.  
 We assume the effective speed of the particle in the analytical theory to vary linearly with the local density around it as: $v(\rho (t)) = v_{0} (1 - \lambda \rho (t))$, where the value of the fitting constant $\lambda$ is derived later from numerical data. We further calculate the mean square displacement of such a particle by solving the governing equations \ref{eqn_pos} and \ref{eqn_theta}. $\boldsymbol{r}(t) = \int_{0}^{t} v(\rho(t')){\bf \hat{n}}(t') dt'$ and mean square displacement
    $\Delta_{ef}(t) = \langle \boldsymbol{r}(t) \cdot \boldsymbol{r}(t) \rangle$
    \begin{gather*}
    \Delta_{ef} (t) = \langle \int_{0}^{t} \int_{0}^{t} v(\rho(t_1)){\bf \hat{n}}_1 \cdot v(\rho(t_2)){\bf \hat{n}}_2dt_1 dt_2 \rangle
\end{gather*}
Using the exact functional form of the self-propulsion speed of the particle: $v(\rho (t)) = v_{0} (1 - \lambda \rho (t))$, and expressing local density in terms of fluctuations as $\rho (t) = \rho_0 + \delta \rho (t)$ gives $v(\rho (t)) = v_{0} (1 - \lambda \rho_0 -\lambda \delta \rho (t))$. Now, we write the mean squared displacement in terms of local density fluctuations and use its temporal mean and auto-correlation properties as a stationary stochastic variable from \ref{rho_mean} and \ref{rho_corr}:
\begin{multline}
    \Delta_{ef}(t) = {v^2_{0}}(1 - \lambda \rho_0)^2 \underbrace{\int_{0}^{t} \int_{0}^{t}  \langle{\bf\hat{n}}_1\cdot{\bf \hat{n}}_2\rangle dt_1 dt_2}_{(a)} \\
    + {v^2_{0}} \lambda^2 \underbrace{\int_{0}^{t} \int_{0}^{t}  \mathcal{C}_\rho (t_2 - t_1) \langle{\bf\hat{n}}_1\cdot{\bf\hat{n}}_2\rangle dt_1 dt_2}_{(b)} 
    \label{MSD}
\end{multline}
Here, $\langle{\bf\hat{n}}_1\cdot{\bf\hat{n}}_2 \rangle = \langle cos\theta_1 cos \theta_2 + sin \theta_1 sin \theta_2 \rangle$ is the orientation correlation, where $\theta_1$ and $\theta_2$ are orientations of the particle at two time instants $t_1$ and $t_2$ respectively. To evaluate the integrals in terms $(a)$ and $(b)$, we put $\mathcal{C}_\rho (t_2 - t_1) = \sigma^2_\rho \exp(-(t_2 - t_1)/\tau)$ from equation \ref{rho_corr} and calculate $\langle{\bf\hat{n}}_1\cdot{\bf\hat{n}}_2 \rangle$ by solving for $\theta (t)$ from the equation \ref{eqn_theta}. The orientation vector of the particle undergoes Brownian motion so the time evolution of its probability distribution is governed by the equation $\frac{\partial}{\partial t} P(\theta, t) = D_{R}\frac{\partial^2}{\partial \theta^2} P (\theta, t)$, where $D_R$ is the rotational diffusion constant and $\theta \in (-\pi, \pi]$. Assuming the particle starts along a direction with orientation $\theta = \theta_0$, the solution of the partial differential equation for $P(\theta, t)$ is given by: \cite{Chubynsky2017}
\begin{equation}
    P(\theta_t | \theta_0, t) = \frac{1}{2\pi}\sum_{m = -\infty}^{+\infty} e^{i m (\theta_t - \theta_0)}e^{-m^2 D_R t}
\end{equation} 
and can be exactly evaluated in terms of a complex elliptic function as:
$
    P(\theta_t | \theta_0, t) = \frac{1}{2\pi}
    \vartheta_3 \Big
    (\frac{\theta_t - \theta_0}{2}, e^{-D_Rt}\Big)
$, where $\vartheta_3 (z,q) = 1 + 2\sum_{m= 1}^{\infty} q^{m^{2}} cos(2mz)$.
Using the probability distribution function $P(\theta , t)$, the orientation correlation $\langle {\bf\hat{n}}_1\cdot{\bf\hat{n}}_2 \rangle = \langle cos\theta_1 cos\theta_2 + \sin\theta_1 sin\theta_2 \rangle $ can be procedurally calculated. First, we calculate the expectation value of cosine and sine of $\theta_t$ starting from orientation $\theta_0$ at initial time $t_0$:
\begin{equation}
    \langle cos \theta_t \rangle_{\theta_0} = \int_{-\pi}^{\pi} cos\theta_t P (\theta_t|\theta_0, t) d\theta_t = e^{-D_R t} cos\theta_0
    \label{exp_cos}
\end{equation}
\begin{equation}
    \langle sin \theta_t \rangle_{\theta_0} = \int_{-\pi}^{\pi} sin\theta_t P (\theta_t|\theta_0, t) d\theta_t = e^{-D_R t} sin\theta_0
    \label{exp_sin}
\end{equation}
Now we calculate the correlation of the cosine and sine of particle orientations at time $t_1$ and $t_2$ using the expectation values calculated in \ref{exp_cos} and \ref{exp_sin}.
\begin{equation}
  \langle cos\theta_1 cos \theta_2 \rangle = \int \int cos\theta_1 cos\theta_2 P(\theta_1,t_1;\theta_2,t_2) d\theta_1 d\theta_2
  \label{acf}
\end{equation}
The joint probability distribution can be expressed as probability of transition from orientation $\theta_1$ to $\theta_2$, given $t_2>t_1$, as $P(\theta_1,t_1;\theta_2,t_2) = P(\theta_2,t_2 | \theta_1, t_1)P(\theta_1, t_1)$. Using this relation, we rewrite the correlation function in equation \ref{acf} as:
$ \langle cos\theta_1 cos \theta_2 \rangle = 
  \int cos\theta_1 P(\theta_1,t_1) \Big(\int cos\theta_2 P(\theta_2,t_2|\theta_1,t_1) d\theta_2 \Big) d\theta_1$. Using equation \ref{exp_cos} and definition of the propagator $P(\theta_t|\theta_0, t)$ we further simplify the expression in two steps. First simplifying the inner integral $\int_{-\pi}^{\pi} cos\theta_2 P(\theta_2,t_2|\theta_1,t_1) d\theta_2 = \langle cos \theta_2 \rangle_{\theta_1} = e^{-D_R (t_2 - t_1)} cos \theta_1$, which gives:  
\begin{equation*}
    \langle cos\theta_1 cos \theta_2 \rangle_{t_2>t_1} =
  e^{-D_R (t_2 - t_1)}\int_{-\pi}^{\pi} cos^2\theta_1 P(\theta_1|\theta_0,t_1) d\theta_1
\end{equation*}
Next, simplifying $\int_{-\pi}^{\pi} cos^2\theta_1 P(\theta_1|\theta_0,t_1) d\theta_1 = \langle cos^2 \theta_1\rangle_{\theta_0}$
\begin{equation}
    \langle cos\theta_1 cos \theta_2 \rangle_{t_2 > t_1} =
  \frac{1}{2}e^{-D_R (t_2 - t_1)} (1+ e^{-4 D_R t_1}cos\theta_0)
  \label{corr_cos}
\end{equation}
Similarly, the auto-correlation of sine of particle orientation $\theta_t$ is calculated
\begin{equation}
    \langle sin\theta_1 sin \theta_2 \rangle_{t_2 > t_1} =
  \frac{1}{2}e^{-D_R (t_2 - t_1)} (1 - e^{-4D_R t_1}cos\theta_0)
  \label{corr_sin}
\end{equation}

\begin{figure}[H]
\begin{minipage}{0.47\textwidth}
    \includegraphics[width=\linewidth, height=11.0cm, keepaspectratio]{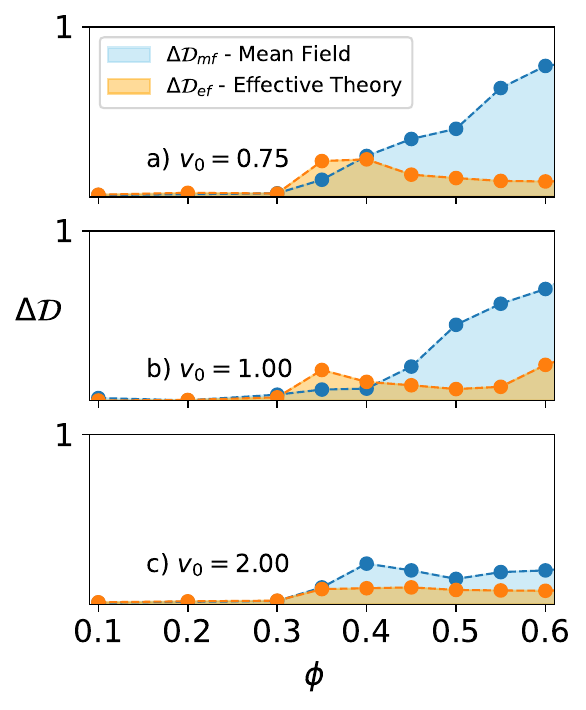}
    \caption{Relative error in the diffusivity value estimated by effective theory (orange) and by mean-field theory (blue). The error is calculated from numerical values of diffusivity $\mathcal{D}_0 (\phi)$ for activity $v_0 = 0.75, 1.0, 2.0$.}
    
    \label{error}
\end{minipage}
\end{figure}

 Following \ref{corr_cos} and \ref{corr_sin} the moment $\langle{\bf \hat{n}}_1\cdot{\bf \hat{n}}_2\rangle$ is:
 \begin{equation}
     \langle{\bf \hat{n}}_1\cdot{\bf \hat{n}}_2\rangle_{t_2 > t_1} = e^{-D_R (t_2 - t_1)}
     \label{corr_orientation}
 \end{equation}
 Hence using correlation functions of particle orientation  $\langle{\bf \hat{n}}_1\cdot{\bf \hat{n}}_2\rangle $ \ref{corr_orientation} and local density fluctuation $\mathcal{C}_{\rho}(t)$ \ref{rho_corr} the terms $(a)$ and $(b)$ in expression \ref{MSD} are evaluated to arrive at an exact expression for the particle's  mean squared displacement $\Delta_{ef} (t)$. We are interested in predicting the effective single-particle transport properties, so from the mean squared displacement we obtain the late-time diffusivity of the particle using the relation $\mathcal{D}_{ef} = \lim_{t \rightarrow \infty} \frac{\Delta_{ef}(t)}{4 t}$
\begin{equation}
    \mathcal{D}_{ef} (\phi) = \frac{{v^2_{0}}}{2 D_R}(1 -  \lambda \rho_0)^2 + \frac{{v^2_{0}}}{2 \beta}\lambda^2 \sigma^2_\rho
    \label{eq:deff}
\end{equation} 
where $\beta = (D_R + \frac{1}{\tau})$. Therefore, our single-particle analytical theory estimates the diffusivity of an active particle in a system with any packing density $\phi$ (dilute or dense) by capturing the steric interactions it experiences as  effective local density fluctuations around a particle. In the dilute regime when density fluctuations are negligible, $\sigma^2_\rho$ and $\tau$ both approach 0 in the bulk limit, and the second term in expression \ref{eq:deff} vanishes assuming the mean-field form. In literature \cite{active_1}, $\mathcal{D}_{mf}(\phi) = \frac{v_0^2}{2 D_R}(1- \lambda_0 \phi)^2$ with $\lambda_0 \approx 0.9$, so we set our fitting parameter to $\lambda = \lambda_0 (\frac{\phi}{\rho_0})$. The slope $\frac{\phi}{\rho_0}$ is derived from the plot \ref{rho_mean_var} (a) for $\phi < \phi_c$, which gives $\lambda \approx 1.2$. In the dense regime the mean-field theory is previously found to fail as with the onset of MIPS density fluctuations become significant. However, the expression $\mathcal{D}_{ef}(\phi)$ fed with local density fluctuation parameters $\rho_0 (\phi), \sigma^2_\rho(\phi) $ and $\tau(\phi)$ estimates the particle's effective diffusivity. Fitting constant is again set to $\lambda = \lambda_0 (\frac{\phi}{\rho_0})$, with a different slope $\frac{\phi}{\rho_0}$ as seen in Fig. \ref{rho_mean_var} (a), which gives $\lambda \approx 0.76$. We test our effective theory and also compare it with mean-field theory using numerical values of diffusivity $\mathcal{D}_{0}(\phi)$ as the baseline. $\mathcal{D}_{0}(\phi) = \lim_{t \rightarrow \infty} \frac{\Delta_0 (t)}{4t}$, where $\Delta_0(t)$ is calculated by numerically simulating equations \ref{eqn_pos} and \ref{eqn_theta}. 
Comparison is drawn by computing relative error from the numerical values:  $\Delta \mathcal{D}_{ef}(\phi) = \frac{\mathcal{D}_{ef}(\phi)-\mathcal{D}_{0}(\phi)}{\mathcal{D}_{0}(\phi)}$ and $\Delta \mathcal{D}_{mf}(\phi) = \frac{\mathcal{D}_{mf}(\phi)-\mathcal{D}_{0}(\phi)}{\mathcal{D}_{0}(\phi)}$. From Fig \ref{error} it is evident that the effective theory is especially more accurate than the mean field for values of system packing density $\phi$ and activity $v_0$ for which there is phase separation in the system, as our theory accounts for the dynamic formation and dissolution of macroscopic clusters in the non-equilibrium steady state. The slight bias observed at high densities arises from the linear speed ansatz; incorporating higher-order density terms could improve accuracy in this regime.  
%%%%%%%%%%%%%%%%%%%%%%%%Discussion%%%%%%%%%%%%%%%%%%%%%%%%%%%%%%
%%%%%%%%%%%%%%%%%%%%%%%%%%%%%%%%%%%%%%%%%%%%%%%%%%%%%%%%%%%%%%%%
 \textit{Discussion-} We theoretically developed a model for the non-equilibrium steady state of active Brownian particles by analyzing local density fluctuations. In the low global packing density regime, these fluctuations are small, leading to a nearly constant local density around a single particle. Consequently, the particle's effective dynamics can be accurately described using mean-field theory, which assumes a uniform density background. At high global packing densities, significant density fluctuations arise, necessitating their inclusion in the effective single-particle dynamics. We numerically computed the mean and temporal autocorrelation of the stochastic local density variable and used these properties to derive effective Langevin equations of motion. By analytically solving these equations, we obtained an exact expression for diffusivity. A key result of our work is the alignment of these numerical findings with the well-established first-order-like phase transition in the system, characterized by the emergence of a dynamic clustered liquid phase coexisting with a surrounding gas phase. More remarkably, our effective single-particle analytical theory remains valid across the entire range of packing densities, whereas mean-field theory is accurate only at low densities. Because our approach explicitly incorporates density fluctuations and their temporal correlations, it remains robust even in the high-density regime beyond the onset of motility-induced phase separation (MIPS), where prominent cluster formation occurs.

Previous studies have either utilized microscopic numerical simulations \cite{active_1,mips_3, mips_4,mips_5,  mips_7,mips_8, mips_9}  or employed effective coarse-grained models to examine the many-particle effect on the local density in the non-equilibrium steady-state of active matter systems.\cite{wittkowski2014scalar,cates2015motility}. However, in our current work, interactions with other particles are accounted for through an effective single-particle model. The key feature of this model is the time dependence of the local density surrounding each particle, characterized by a correlation time. Notably, this correlation time can be experimentally measured in active matter systems such as quorum-sensing microswimmers \cite{quorum_sensing}. Once determined, it allows us to replace the many-body interacting system with our effective single-particle model. \\
This simplification promises to streamline the analysis and understanding of complex interacting active particle systems. The current study is focused to two-dimensions but the effective theory can be easily generalized to higher dimensions \cite{jayam_paper}.

\acknowledgments
The support and the resources provided by PARAM Shivay Facility under the National Supercomputing Mission, Government of India at the Indian Institute of Technology, Varanasi are gratefully acknowledged by all authors. S.M. thanks DST-SERB India, ECR/2017/000659, CRG/2021/006945 and MTR/2021/000438 for financial support. J.J, P.K.M and S. M. thank the Centre for Computing and Information Services at IIT (BHU), Varanasi.

\end{document}